\documentclass[aps,prl,twocolumn,superscriptaddress,showpacs,showkeys,amsmath,amssymb,floatfix]{revtex4}%preprint <-> twocolumn
\usepackage{booktabs,graphicx,mathrsfs,verbatim}
\usepackage[colorlinks=true,citecolor=blue,filecolor=blue,linkcolor=blue,urlcolor=blue,pdftex]{hyperref}
\usepackage[usenames,dvipsnames]{color}
\usepackage{ulem}
\usepackage{color}

\newcommand{\assergi}{\affiliation{INFN, Laboratori Nazionali del Gran Sasso, Assergi, 67100, Italy}}
\newcommand{\bern}{\affiliation{Albert Einstein Center for Fundamental Physics, University of Bern, Sidlerstrasse 5, 3012 Bern, Switzerland}}
\newcommand{\bologna}{\affiliation{University of Bologna and INFN-Bologna, Bologna, Italy}}
\newcommand{\columbia}{\affiliation{Physics Department, Columbia University, New York, NY 10027, USA}}
\newcommand{\coimbra}{\affiliation{Department of Physics, University of Coimbra, R. Larga, 3004-516, Coimbra, Portugal}}
\newcommand{\heidelberg}{\affiliation{Max-Planck-Institut f\"ur Kernphysik, Saupfercheckweg 1, 69117 Heidelberg, Germany}}
\newcommand{\houston}{\affiliation{Department of Physics and Astronomy, Rice University, Houston, TX 77005 - 1892, USA}}

\newcommand{\losangeles}{\affiliation{Physics \& Astronomy Department, University of California, Los Angeles, USA}}
\newcommand{\mainz}{\affiliation{Institut f\"ur Physik, Johannes Gutenberg-Universit\"at Mainz, 55099 Mainz, Germany}}
\newcommand{\munster}{\affiliation{Institut f\"ur Kernphysik, Wilhelms-Universit\"at M\"unster, 48149 M\"unster, Germany}}
\newcommand{\nikhef}{\affiliation{Nikhef  and the University of Amsterdam, Science Park, Amsterdam, Netherlands}}
\newcommand{\purdue}{\affiliation{Department of Physics, Purdue University, West Lafayette, IN 47907, USA}}
\newcommand{\shanghai}{\affiliation{Department of Physics, Shanghai Jiao Tong University, Shanghai, 200240, China}}
\newcommand{\subatech}{\affiliation{SUBATECH, Ecole des Mines de Nantes, CNRS/In2p3, Universit\'e de Nantes, 44307 Nantes, France}}
\newcommand{\torino}{\affiliation{INFN-Torino and Osservatorio Astrofisico di Torino, 10100 Torino, Italy}}
\newcommand{\weizmann}{\affiliation{Department of Particle Physics and Astrophysics, Weizmann Institute of Science, 76100 Rehovot, Israel}}
\newcommand{\zurich}{\affiliation{Physics Institute, University of Z\"{u}rich, Winterthurerstr. 190, CH-8057, Switzerland}}

\begin{document}
\title{Limits on spin-dependent WIMP-nucleon cross sections from 225 live days of XENON100 data}

\author{E.~Aprile}\columbia 
\author{M.~Alfonsi}\nikhef
\author{K.~Arisaka}\losangeles
\author{F.~Arneodo}\assergi
\author{C.~Balan}\coimbra
\author{L.~Baudis}
\email{laura.baudis@physik.uzh.ch}
\zurich
\author{B.~Bauermeister}\mainz
\author{A.~Behrens}\zurich
\author{P.~Beltrame}\weizmann\losangeles
\author{K.~Bokeloh}\munster
\author{A.~Brown}\purdue
\author{E.~Brown}\munster
\author{G.~Bruno}\assergi
\author{R.~Budnik}\columbia 
\author{J.~M.~R.~Cardoso}\coimbra
\author{W.-T.~Chen}\subatech
\author{B.~Choi}\columbia
\author{A.~P.~Colijn}\nikhef
\author{H.~Contreras}\columbia
\author{J.~P.~Cussonneau}\subatech
\author{M.~P.~Decowski}\nikhef
\author{E.~Duchovni}\weizmann
\author{S.~Fattori}\mainz
\author{A.~D.~Ferella}\assergi\zurich
\author{W.~Fulgione}\torino
\author{F.~Gao}\shanghai
\author{M.~Garbini}\bologna
\author{C.~Ghag}\losangeles
\author{K.-L.~Giboni}\columbia
\author{L.~W.~Goetzke}\columbia
\author{C.~Grignon}\mainz
\author{E.~Gross}\weizmann
\author{W.~Hampel}\heidelberg
\author{F.~Kaether}\heidelberg
\author{A.~Kish}\zurich
\author{J.~Lamblin}\subatech
\author{H.~Landsman}
\email{hagar.landsman@weizmann.ac.il}
\weizmann
\author{R.~F.~Lang}\purdue
\author{M.~Le~Calloch}\subatech
\author{D.~Lellouch}\weizmann
\author{C.~Levy}\munster
\author{K.~E.~Lim}\columbia
\author{Q.~Lin}\shanghai
\author{S.~Lindemann}\heidelberg
\author{M.~Lindner}\heidelberg
\author{J.~A.~M.~Lopes}\coimbra
\author{K.~Lung}\losangeles
\author{T.~Marrod\'an~Undagoitia}\heidelberg\zurich
\author{F.~V.~Massoli}\bologna
\author{A.~J.~Melgarejo~Fernandez}\columbia
\author{Y.~Meng}\losangeles
\author{M.~Messina}\columbia
\author{A.~Molinario}\torino
\author{K.~Ni}\shanghai
\author{U.~Oberlack}\mainz
\author{S.~E.~A.~Orrigo}\coimbra
\author{E.~Pantic}\losangeles
\author{R.~Persiani}\bologna
\author{G.~Plante}\columbia
\author{N.~Priel}\weizmann
\author{A.~Rizzo}\columbia
\author{S.~Rosendahl}\munster
\author{J.~M.~F.~dos Santos}\coimbra
\author{G.~Sartorelli}\bologna
\author{J.~Schreiner}\heidelberg
\author{M.~Schumann}\bern\zurich
\author{L.~Scotto~Lavina}\subatech
\author{P.~R.~Scovell}\losangeles
\author{M.~Selvi}\bologna
\author{P.~Shagin}\houston
\author{H.~Simgen}\heidelberg
\author{A.~Teymourian}\losangeles
\author{D.~Thers}\subatech
\author{O.~Vitells}\weizmann
\author{H.~Wang}\losangeles
\author{M.~Weber}\heidelberg
\author{C.~Weinheimer}\munster

\collaboration{The XENON100 Collaboration}\noaffiliation

\begin{abstract}

We present new experimental constraints on the elastic, spin-dependent WIMP-nucleon cross section using recent data from the XENON100 experiment, operated in the Laboratori Nazionali del Gran Sasso in Italy. An analysis of 224.6 live days\,$\times$\,34\,kg of exposure acquired during 2011 and 2012 revealed no excess signal due to axial-vector WIMP interactions with   $^{129}$Xe and $^{131}$Xe nuclei. This leads to the most stringent upper limits on WIMP-neutron cross sections for WIMP masses above 6\,GeV/c$^2$, with a minimum cross section of  3.5$\times 10^{-40}$\,cm$^2$ at a WIMP mass of 45\,GeV/c$^2$, at 90\% confidence level.

\end{abstract}



\maketitle

XENON100 was built to search for hypothetical, weakly interacting massive particles (WIMPs), which could explain the non-baryonic, cold dark matter in our Universe \,\cite{Jungman:1995df}.
Independently of astrophysical and cosmological observations, WIMPs are a consequence of many extensions of the Standard Model of  particle physics, as new, stable or long-lived neutral particles. 
The WIMP dark matter hypothesis is testable by experiment, the most compelling avenue is to directly observe WIMPs scattering off atomic nuclei in ultra-low background terrestrial detectors\,\cite{Goodman:1984dc,Baudis:2012ds}. 
XENON100 is a double-phase xenon time projection chamber operated at the Laboratori Nazionali del Gran Sasso (LNGS) in Italy. A total of 178 low-radioactivity, UV-sensitive photomultiplier tubes detect the prompt  (S1) and proportional (S2) light signals induced by particles interacting in the sensitive volume, containing 62\,kg of ultra-pure liquid xenon. The background level in the energy region of interest for dark matter searches ($<$50\,keV$_{nr}$) is 5.3$\times$10$^{-3}$\,events\,kg$^{-1}$\,d$^{-1}$\,keV$^{-1}$, before discrimination of electronic and nuclear recoils based on their S2/S1-ratio \cite{Aprile:2012nq,Aprile:2011vb}. The instrument is described in \cite{Aprile:2011dd}, the analysis procedure is detailed in \cite{Aprile:2012vw}.

WIMPs in the halo of our Galaxy are expected to be highly non-relativistic \cite{Green:2011bv} and their interactions with nuclei can be characterized in terms of scalar (or spin-independent, SI) and axial-vector (or spin-dependent, SD) couplings \,\cite{Goodman:1984dc,Jungman:1995df}.  In the case of SI interactions, the leading contribution of the scattering is coherent across the nucleus, and roughly scales with $A^2$, where $A$ is the number of nucleons. Our SI result was presented in 
\cite{Aprile:2012nq} and excludes a WIMP-nucleon cross section above $2\times10^{-45}$\,cm$^2$ at a WIMP mass of 55\,GeV$/c^2$ at 90\% confidence level.  Here we use the same data set, with an exposure of 224.6 live days, a fiducial mass of 34\,kg, identical event selection cuts, acceptances, relative scintillation efficiency and background model to derive limits on spin-dependent interactions.

If the WIMP is a spin-1/2 or a spin-1 field, the contributions to the WIMP-nucleus scattering cross section arise from couplings of the WIMP field to the quark axial current. In the case of the lightest neutralino in supersymmetric models for instance, scattering occurs through the exchange of $Z$ bosons or squarks\,\cite{Jungman:1995df}.  To predict actual rates, these fundamental interactions are first translated into interactions with nucleons by evaluating the matrix element  of the quark axial-vector current in a nucleon. 
Finally, the spin components of the nucleons must be added coherently using nuclear wave functions to yield the matrix element for the SD WIMP-nucleus cross section as a function of momentum transfer.  The SD differential WIMP-nucleus cross section as a function of momentum transfer $q$ can be written as \cite{Engel:1992bf}:

\begin{eqnarray}
\frac{d\sigma_{\mathrm{SD}}(q)}{d{q}^2} =  \frac{8 G_F^2}{ (2J+1){v^2}} S_A(q),
\end{eqnarray}
where $G_F$ is the Fermi constant, $v$ is the WIMP speed relative to the target, $J$ is the total angular momentum of the nucleus and $S_A$ is the axial-vector structure function.  In the limit of zero momentum transfer \footnote{At finite momentum transfer, or when WIMP couplings to two nucleons are included \cite{Menendez:2012tm}, the
  neutron-only coupling case implies also coupling to protons and vice versa.} the structure function reduces to the form \cite{Ressell:1993qm}:

\begin{eqnarray}\label{eq:sij}
S_A(0) = \frac{(2J+1)(J+1)}{\pi J} [a_p\langle S_p\rangle + a_n\langle S_n\rangle ]^2,
\end{eqnarray}
where $\langle S_{p,n}\rangle=\langle J|\hat{S}_{p,n}|J \rangle$ are
the expectation values of the total proton and neutron spin operators in the nucleus, and the effective WIMP couplings to protons and neutrons are defined in terms of the isoscalar $a_0 = a_p+a_n$ and isovector $a_1 = a_p-a_n$ couplings.

WIMPs will thus couple to the total angular momentum of a nucleus and only nuclei with an odd number of protons or/and neutrons will yield a significant sensitivity to this channel.  Natural xenon contains two non-zero spin isotopes, $^{129}$Xe (spin-1/2) and $^{131}$Xe (spin-3/2), with an abundance of $26.4\%$ and $21.2\%$, respectively.
In XENON100, the isotopic abundances of $^{129}$Xe  and $^{131}$Xe are changed to 26.2\% and 21.8\%, respectively.

To compare results from different target materials, a common practice is to report the cross section for the interaction with a single nucleon  ($\sigma_p$, $\sigma_n$) \,\cite{Tovey:2000mm,Giuliani:2004uk,Savage:2004fn}.  Assuming that WIMPs couple predominantly to protons ($a_n=0$) or neutrons ($a_p=0$), the WIMP-nucleon cross section becomes:

\begin{eqnarray}
\sigma_{p,n}(q)=\frac{3}{4}\frac{{\mu_{p,n}^2}}{\mu_{A}^2} \frac{2J+1}{\pi}\frac{ \sigma_{\mathrm{SD}}(q)}{{S_A^{a_0=\pm a_1} (q)}} ,
\label{eq:sigma_Anp} 
\end{eqnarray}
where $\sigma_{SD}$ is the total WIMP-nucleus cross section,  $\mu_{A}$ and $\mu_{p,n}$ are the WIMP-nucleus and WIMP-nucleon reduced masses, respectively. 

\begin{figure}[!h]
%\centering
\includegraphics[width=1\columnwidth]{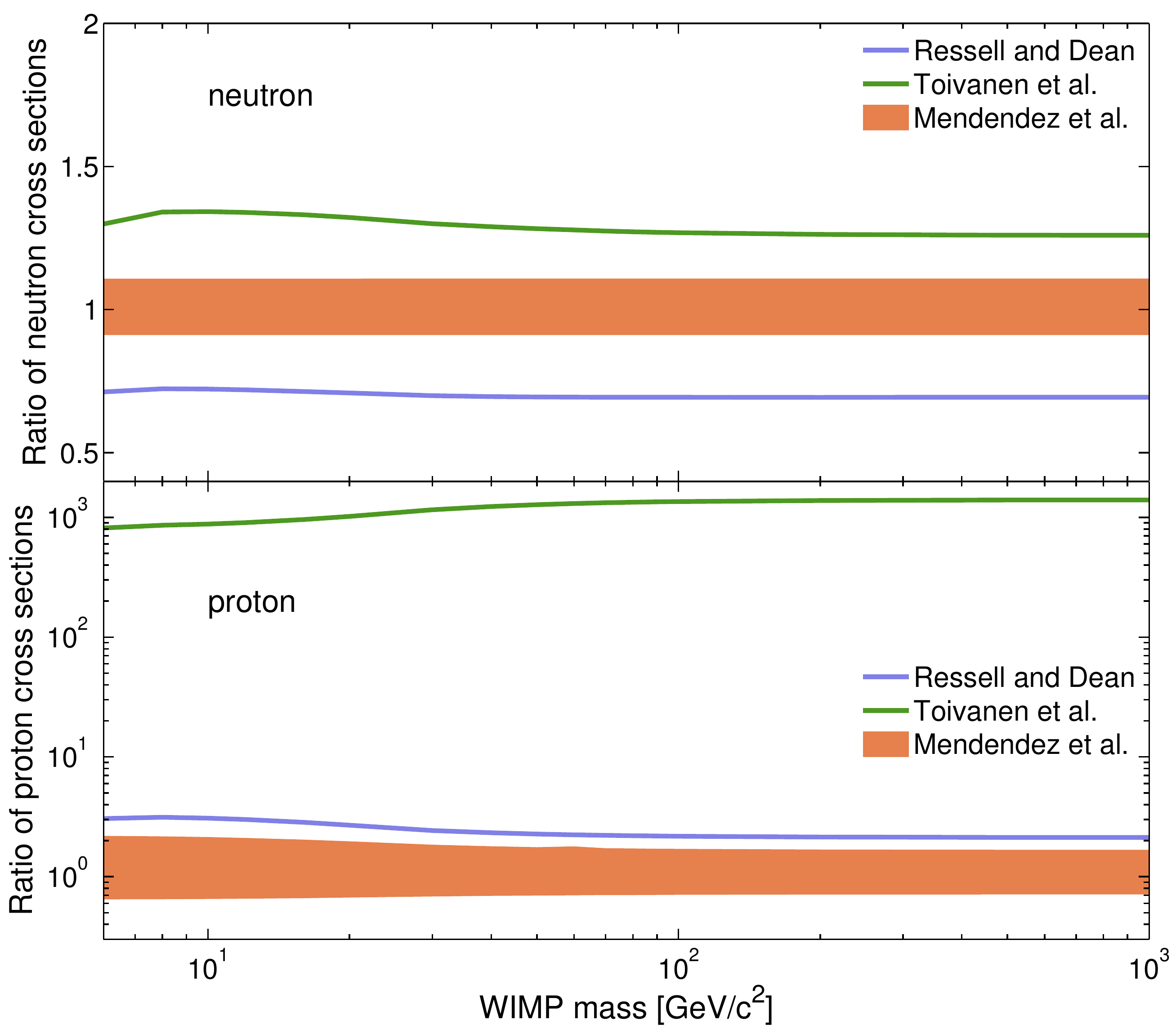}
\caption{The ratio of upper limits calculated with the Ressell and Dean\,\cite{Ressell:1997kx} and  the Toivanen {\it et al.} \,\cite{Toivanen:2009zza} results for the structure functions to the ones obtained using  Menendez {\it et al.} \cite{Menendez:2012tm} for the case of neutron (top) and proton (bottom) couplings, along with the theoretical uncertainty band due to chiral two-body currents  \cite{Menendez:2012tm}.}
\label{fig:theo_errors}
\end{figure}

Calculations of the structure functions $S_A(q)$ are traditionally based on the nuclear shell model, but differ in the effective nucleon-nucleon interactions and in the valence space and truncation used for the computation.   For xenon as a WIMP target material, we consider three large-scale shell-model calculations: by Ressell and Dean~\cite{Ressell:1997kx} with the  Bonn-A~\cite{HjorthJensen:1995ap} two-nucleon potential, by Toivanen {\it et al.}~\cite{Toivanen:2009zza},  using the CD-Bonn potential \cite{HjorthJensen:1995ap}, and the recent results by Menendez {\it et  al.} \cite{Menendez:2012tm},  using state-of-the-art valence shell interactions \cite{Caurier:2007wq, Menendez:2008jp} and less severe truncations of the valence space. Menendez {\it et al.} \cite{Menendez:2012tm} also use for the first time chiral effective field theory (EFT) currents \cite{Park:2002yp} to determine the couplings of WIMPs to nucleons. The currents for spin-dependent scattering are derived at the one-body level and  the leading long-range two-nucleon currents are included, resulting in a reduction of the isovector part of the one-body axial-vector WIMP currents  \cite{Menendez:2012tm}. The resulting chiral EFT currents are then used to calculate the structure functions for the WIMP-xenon scattering. Theoretical errors due to nuclear uncertainties can be provided when chiral two-body currents are included \cite{Menendez:2012tm}. We show their effect in Figure \ref{fig:theo_errors}. The shell-model calculations are based on the largest many-body spaces accessible with nuclear interactions, also used to calculate double-beta decay matrix  elements for nuclei up to $^{136}$Xe  and to study nuclear structures \cite{Caurier:2007wq, Menendez:2008jp, Menendez:2011qq}.

The new calculations by Menendez {\it et al.} \cite{Menendez:2012tm} yield a far superior agreement between calculated and measured spectra of the $^{129}$Xe and $^{131}$Xe nuclei, both in energy and in the ordering of the nuclear levels, compared to older \cite{Toivanen:2009zza} results.  The values for ${\langle S_{p,n}\rangle}$ are close to those of Ressell and Dean~\cite{Ressell:1997kx}, but quite different from the results of Toivanen {\it et al.}~\cite{Toivanen:2009zza}, as summarized in Table \ref{tab:matrix_elements}.   We thus use the Menendez {\it et al.} \cite{Menendez:2012tm} structure functions for our benchmark upper limits on WIMP-neutron and WIMP-proton cross sections. We also provide a comparison to the limits obtained when using the calculations by  Ressell and Dean~\cite{Ressell:1997kx}  and Toivanen {\it et al.}~\cite{Toivanen:2009zza}. In all cases, $|{\langle S_{n}\rangle}| \gg |{\langle S_{p}\rangle}|$, as expected for the two xenon nuclei with an odd number of neutrons and an even number of protons. 

\begin{figure}[!h]
\centering
\includegraphics[width=1\columnwidth]{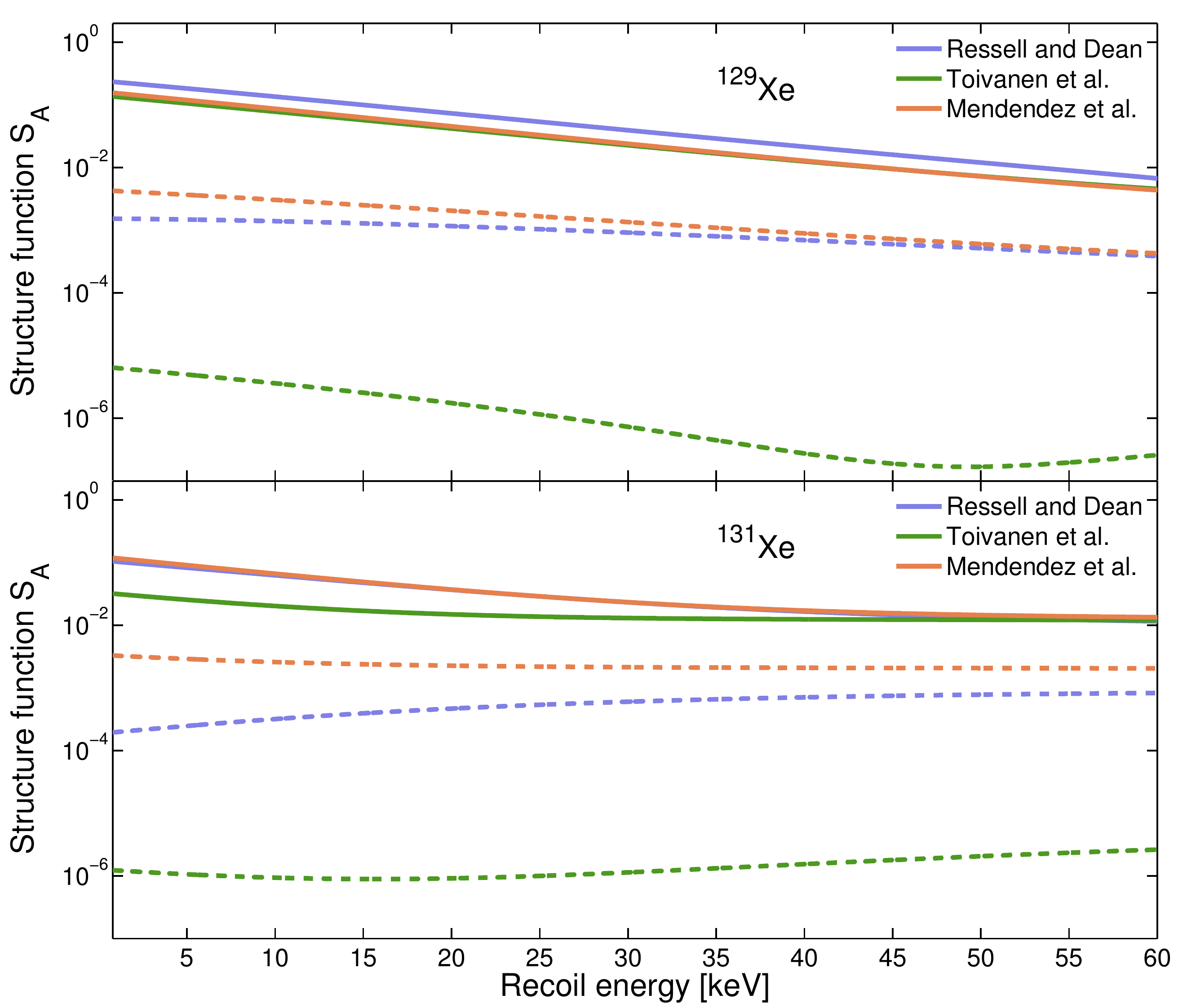}
\caption{Structure functions for $^{129}$Xe (top) and $^{131}$Xe (bottom) for the case of neutron (plain) and proton (dashed) couplings,  as a function of recoil energy using the calculations of  Ressell and Dean\,\cite{Ressell:1997kx}, Toivanen {\it et al.}\,\cite{Toivanen:2009zza} and Menendez {\it et al.} \cite{Menendez:2012tm}. The difference is most significant in the case of the proton coupling for the Toivanen {\it et al.} results.} 
\label{fig:structure_fct}
\end{figure}

\begin{table*}
\caption{ Parameters of the xenon isotopes used in this analysis: nuclear total angular momentum and parity of the ground state, $J^P$, predicted expectation values of the total proton and neutron spin operators in the nucleus $\langle S_{n,p}\rangle$ by the Ressell and Dean (Bonn~A potential)~\cite{Ressell:1997kx}, Toivanen {\it et al.}  (Bonn~CD  potential)~\cite{Toivanen:2009zza} and  Menendez {\it et al.}  (state-of-the art valence shell interactions) \cite{Menendez:2012tm}  calculations.} 
\label{tab:matrix_elements}
\begin{tabular}{l|c|cc|cc|cc}
\multicolumn{2}{c}{}&\multicolumn{2}{c|}{Ressell and Dean~\cite{Ressell:1997kx}}&\multicolumn{2}{c|}{Toivanen {\it et al.}~\cite{Toivanen:2009zza}}&\multicolumn{2}{c}{ Menendez {\it et al.} \cite{Menendez:2012tm} }\\
\hline
Nucleus & $J^P$ & $\langle S_n \rangle$ & $\langle S_p \rangle$ &   $\langle S_n \rangle$ & $\langle S_p \rangle$&  $\langle S_n \rangle$ & $\langle S_p \rangle$\\
\hline
$^{129}$Xe & $\left(\frac{1}{2}\right)_{g.s.}^{+}$ &  $\;\;0.359$ & $\;\;0.028$  & $\;\;0.273$ & $-0.0019$  & $\;\;0.329$ & $\;\;0.010$ \\
$^{131}$Xe & $\left(\frac{3}{2}\right)_{g.s.}^{+}$ &  $-0.227$ & $-0.009$  & $-0.125$ & $-0.00069$ & $-0.272$ & $-0.009$ \\
\end{tabular}
\end{table*}

Figure~\ref{fig:structure_fct} shows the structure functions $S_A(q)$ obtained from the three calculations for pure neutron and pure proton couplings as a function of nuclear recoil energy.  For the neutron coupling case, for which xenon has the highest sensitivity, the functions are rather similar.  For the proton coupling case, the structure function by  Toivanen {\it et al.}~\cite{Toivanen:2009zza} differs significantly from the other two results. Table~\ref{tab:matrix_elements}  summarizes the expectation values of the total proton and neutron spin operators in the nucleus for $^{129}$Xe  and $^{131}$Xe in the zero momentum transfer limit.

\begin{figure}[!h]
\centering
\includegraphics[width=1\columnwidth]{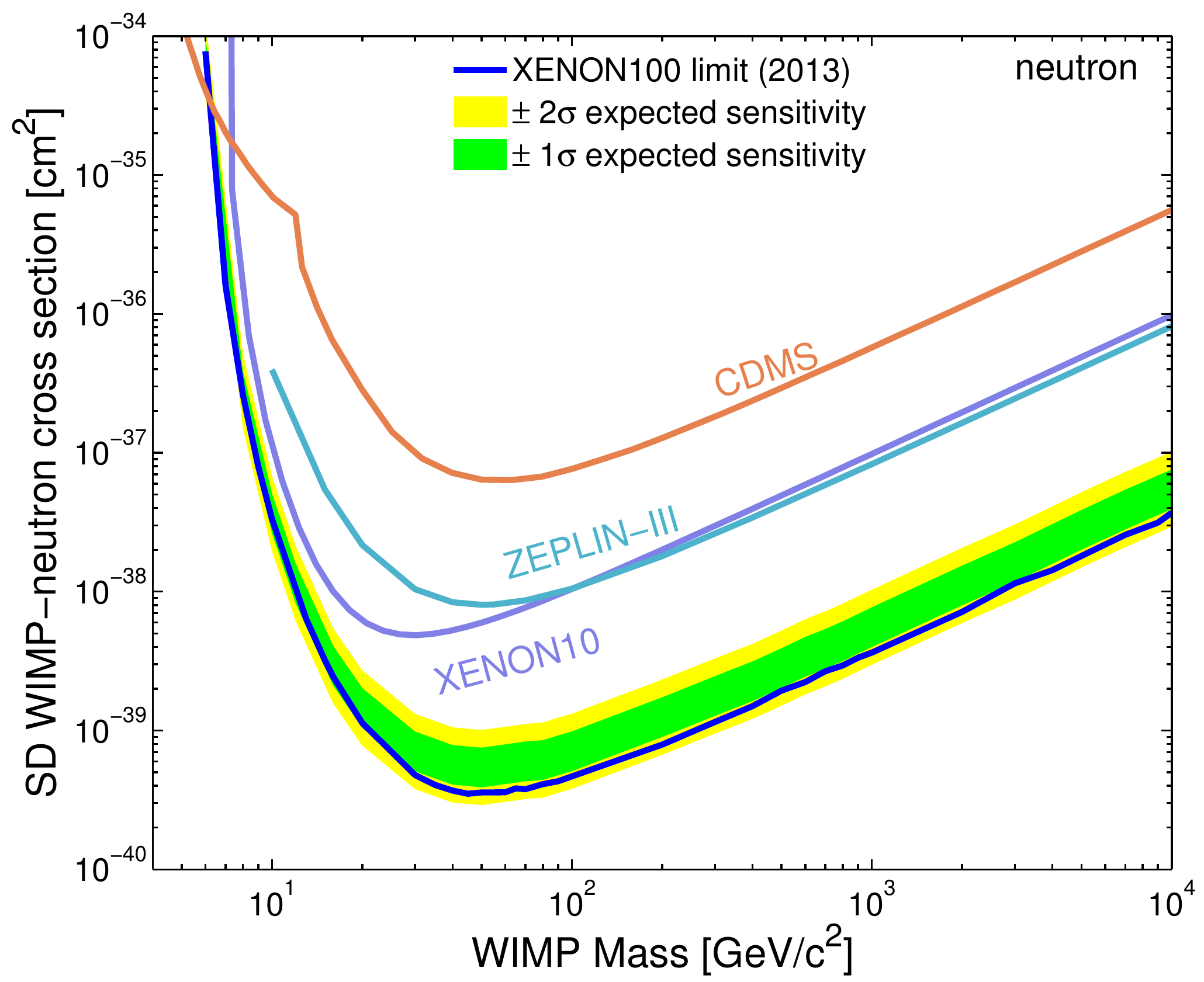}
\includegraphics[width=1\columnwidth]{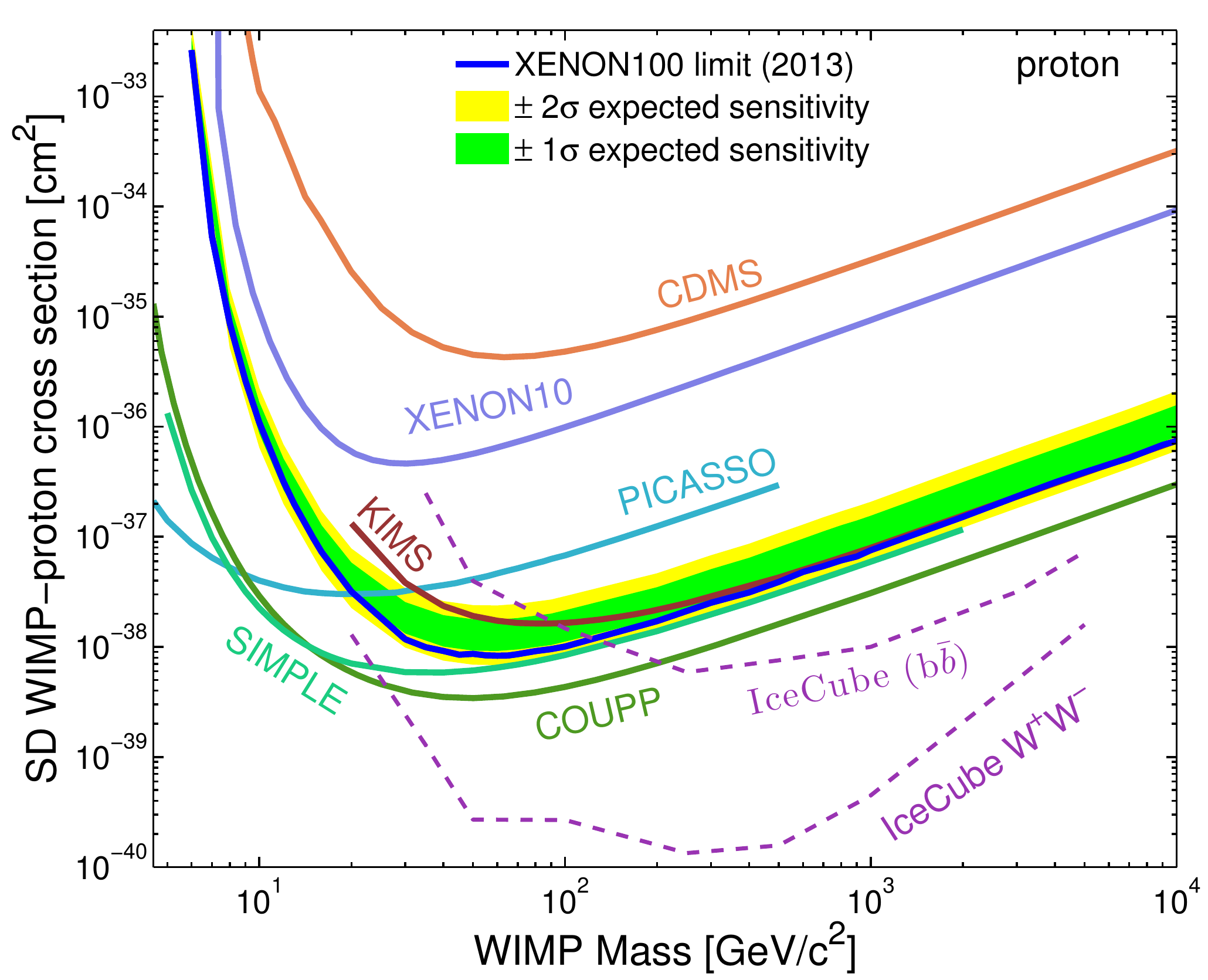}
\caption{XENON100 90\%\,C.L. upper limits on the WIMP SD cross section on neutrons (top) and protons (bottom) using Menendez {\it et al.} \cite{Menendez:2012tm}. The 1$\sigma$ (2$\sigma$) uncertainty on the expected sensitivity of this run is show as a green (yellow) band.
Also shown are results from XENON10\,\cite{Angle:2008we} (using Ressel and Dean \,\cite{Ressell:1997kx}), CDMS\,\cite{Ahmed2009,Ahmed:2010wy}, ZEPLIN-III\,\cite{Akimov:2011tj} (using Toivanen {\it et al.}   \,\cite{Toivanen:2009zza}), PICASSO\,\cite{Archambault:2012pm} , COUPP\,\cite{Behnke:2012}, SIMPLE\,\cite{Felizardo:2011uw}, KIMS\,\cite{Kim:2012rza}, IceCube\,\cite{Aartsen:2012ef} in the hard ($W^+$$W^-$, $\tau^+\tau^-$ for WIMP masses $<$80.4\,GeV/c$^2$) and soft ($b\bar{b}$) annihilation channels. }
\label{fig:sd_limits}
\end{figure}

Constraints on the spin-dependent WIMP-nucleon cross sections are calculated  using the Profile Likelihood approach described in~\cite{Aprile:2011hx}.
Systematic uncertainties in the energy scale and in the background expectation are taken into account when constructing the Profile Likelihood model and are reflected in the actual limit. It is given at 90\% C.L. after taking into account statistical downward fluctuations in the background. We assume that the dark matter is distributed in an isothermal halo with a truncated Maxwellian velocity distribution with a local circular speed of $v_c = 220$\,km/s, galactic escape velocity $v_{\textrm{esc}} = 544$\,km/s and a local density of $\rho= 0.3$\,GeV\,cm$^{-3}$\,\cite{Green:2011bv}. 

The resulting upper limits from XENON100, along with results from other experiments,  are shown in Figure~\ref{fig:sd_limits} for neutron couplings (top panel) and proton couplings (lower panel).  The 1\,$\sigma$ (2\,$\sigma$) uncertainty on the sensitivity of this run, namely the expected limit in absence of a signal above the background, is shown as  a green (yellow) band  in Figure~\ref{fig:sd_limits}.  The impact on these limits when using the Toivanen {\it et al.} and the Ressell and Dean calculations are shown in Figure \ref{fig:theo_errors}.  XENON100 provides the most stringent limits for pure neutron couplings for WIMP masses above 6\,GeV/$c^2$, excluding previously unexplored regions in the allowed parameter space. The minimum 
WIMP-neutron cross section is 3.5$\times$10$^{-40}$\,cm$^2$ at a WIMP mass of 45\,GeV/c$^2$, using  Menendez {\it et al.} \cite{Menendez:2012tm}.  It changes to  2.5$\times$10$^{-40}$cm$^2$  and 4.5$\times$10$^{-40}$cm$^2$  when using Ressell and Dean and Toivanen {\it et al.}, respectively.
The sensitivity to proton couplings (Figure~\ref{fig:sd_limits}, bottom panel) is much weaker  because, as detailed above,  both $^{129}$Xe and $^{131}$Xe have an unpaired neutron but an even number of protons,  thus $|\langle S_{p}\rangle |\ll |\langle S_{n}\rangle|$ (see table~\ref{tab:matrix_elements}).  Upper limits from other direct and indirect detection experiments are shown for comparison. 

In conclusion, we have analyzed data from 224.6 live days\,$\times$\,34\,kg exposure acquired by XENON100 during 13 months of operation in 2011/2012 for SD WIMP interactions. We saw no evidence for a dark matter signal and have obtained new experimental upper limits on the spin-dependent WIMP-nucleon cross section.   For our limits, we use the new calculations by Menendez {\it et al.} \cite{Menendez:2012tm}, where the WIMP couplings to nucleons are derived using chiral EFT currents and which yield a good agreement between the calculated and measured energy spectra of the $^{129}$Xe and $^{131}$Xe nuclei.
We note that the interpretation of the results in terms of SD pure-proton cross section strongly depends on the used nuclear model. However, regardless of the nuclear model, we obtain the most stringent limits to date on spin-dependent WIMP-neutron couplings for WIMP masses above  6\,GeV$/c^2$ at 90\% C.L.

% Acknowledgements

We gratefully acknowledge support from NSF, DOE, SNF, UZH, FCT, INFN, R\'egion des Pays de la Loire, STCSM, NSFC, DFG, Stichting voor Fundamenteel Onderzoek der Materie (FOM), the Max Planck Society and the Weizmann Institute of Science. We thank Achim Schwenk and Javier Menendez for many helpful discussions and for providing their structure functions in numerical form.  We thank J. Suhonen for providing us the numerical data for Figure \ref{fig:theo_errors} and M. Pitt  (WIS) for his contribution. We are grateful to LNGS for hosting and supporting XENON100.

\bibliographystyle{apsrev}

\bibliography{SD_XENON100_v11}

\end{document}